# C# Traceability System

MICHAEL KERNAHAN, MIRIAM CAPRETZ, LUIZ CAPRETZ
Department of Electrical and Computer Engineering
University of Western Ontario
London, Ontario
CANADA
mkernaha@uwo.ca, mcapretz@eng.uwo.ca, lcapretz@eng.uwo.ca

*Abstract:*       Traceability information is a valuable asset that software development teams can leverage to minimise their risk during production and maintenance of software projects.  When maintainers are added to a software project post-production, they have to learn the system from scratch and understand its dynamics before they can begin making appropriate modifications to the source code.  The system outlined in this paper extracts traceability information directly from the source code of C# projects, and presents it in such a way that it can be easily used to understand the logic and validate changes to the system.

## 1  Introduction

Maintenance is an inescapable part of the software lifecycle process.  Many organizations are realizing the importance of maintenance efforts to ensure the success of a software project.  Maintenance efforts have become part of the overall software lifecycle process, with predetermined timelines and budgets. The acceptance of maintenance as an inevitable requirement has forced companies to examine their policies for hiring and training both their developers and their maintainers.  Companies still have one major gripe with maintenance however: it means that they are losing money since the company rarely is able to generate income from maintenance efforts. For this reason, companies are interested in finding and exploiting ways that they can reduce maintenance effort and thus increase the profit margin on a software project.

One of the major hurdles of software maintenance is for the maintainers to actually understand the complete picture of what the original software does and how it does it.  Without this thorough understanding, all but the simplest maintenance tasks can easily create more problems then they solve.  Not only is it hard for the maintainer to determine if the fix has actually solved the problem, but it is almost impossible for the maintainer to ensure that no new bugs have been unsuspectingly added to the code.  So the initial goal of maintenance efforts should be to read the technical documentation and try to understand the goals of the software and how all of the components fit together in the system.  Unfortunately, as many maintainers can attest, technical documentation for software projects is usually lacking in depth and clarity.

Being able to visualize the traceability information for a software project in object-oriented programming develops with intimate knowledge of the system.  Usually, the programmers who originally wrote the software understand what triggers events, and what handles those events.  They also know the major variables within the major classes, and the implications of making changes to these variables.  A maintainer or programmer new to the project will have a difficult time visualizing all of these relationships.  This information can be very helpful in reducing improper usage of variables, and reducing duplicate effort.

This paper proposes a method for pulling out key information from C# source code directly.  By removing the human factor from knowledge extraction from source code, it is hoped to produce useful information independently of coding style, comments, and any documentation that may have been produced.  This method tokenizes the source code and creates an eXtensible Mark-up Language (XML) file that represents all of the *Namespaces*,

*Classes*, *Methods*, *Constructors*, *Variables*, *Events*, *Properties* and *Delegates* in the source file(s). This XML file is then parsed and utilised to create a database that can be used in a graphical user interface to view the traceability information for the code. In the remainder of this paper, Section 2 presents the problems that face maintenance efforts. Section 3 examines some of the research that is being done to address these issues. Section 4 reviews how traceability information is extracted from source code, while Section 5 presents the data model used to represent the information. Section 6 reviews the interface created for this work, Section 7 examines an example of using the traceability information generated from source code, while Section 8 suggests areas for future work in this field.

## 2  Problems for Maintenance

Technical documents written by programmers are usually too short and superficial, or too long and obtuse. On top of this, the documents are usually prepared once the software has been written already. This usually means that details are forgotten, and left out of the documentation. Most programmers on software projects have little or no experience with maintaining a software system that they did not help to create. Because of this, programmers rarely appreciate just how much of their knowledge about a software system is in their heads and not captured in documentation anywhere. This lack of understanding and the usual practice of creating documentation when a project is complete are major factors in the cost of maintenance.

Metrics can be useful for evaluating programs, as well as for trying to understand how a program was written. Managers can use metrics to track the performance of the development team, while programmers can use them to identify problem areas in the code. Metrics lose a lot of value when they are used without the context of how the project is put together and functions.

The use cases of software projects have a tendency to shift during the development stage. These shifts are often not recorded in the documentation, and usually undocumented use cases exist. Maintaining system integrity can be difficult without an understanding of the use cases of the system.

Throughout the development and maintenance stages, software projects are changed almost constantly. If such changes are implemented in an incomplete or inconsistent way, a loss of architectural quality will occur [1]. The lack of available traceability resources is a problem for collaboration during the development and maintenance phases of a software lifecycle.

## 3  Current Research

Vestdam and Nørmark [2, 3], with their Elucidative documentation method, attempt to help programmers with documentation during the coding phase. Marks and Wilkie [4] present the OSCAR tool for extracting metrics from software automatically. Qin *et al.* [5] have studied extracting use cases from source code with this in mind.

Research is being done to look for ways to help the programmers maintain documentation throughout the software lifecycle. With ease of use, and more emphasis on the importance of documentation, solid documentation skills may develop in the industry. Unfortunately, there will always be a gap between what information the programmer records, and what the maintainers need.

Work in the field of traceability analysis for software projects has attempted to fill this gap with the information that the maintainers need. Riebisch [1] has begun work on a system to link design requirements to the actual source code. He has pointed out that many CASE tools could support traceability with minor amounts of effort.

Balzer and Deussen [6] have developed a graphical environment for representing the *package*, *class*, *method* and *attribute* levels of abstraction of Java code. Their Hierarchical Net is useful for seeing the tree like structure of a software system, and visually showing what method or class fits where.

DeLucia *et al.* [7] have acknowledged the tremendous time and effort required to produce meaningful traceability information manually or semi-automatically. They have proposed a solution for finding traceability links between software artefacts. In their solution, both the software engineer and the system identify links. These two groups of links are analysed to find *candidate links* and *warning links*, which may need to be added or removed from the system.

## 4  Traceability Extraction

The first task to address when automating knowledge extraction from source code is to understand how the code will be interpreted into tokens. To accomplish this task, a tokenizer developed by the #Develop[1] open source project has been used. This initial tokenization was performed without making any changes to the tokeniser developed for the #Develop communities' tool which is able to convert C# code to VB.Net code [8] and vice versa.

Once the source code had been parsed and tokenized, the next step was to represent this code in an XML format so that it can be used by other tools, as well this one. Unfortunately, because of the complexity of the code, it was not possible to simply use the serialisation capabilities of object-oriented programs. Serialisation is essentially the automatic mapping of objects into binary or XML files. Since serialisation was not possible with the tokenised data model, the #Develop C# to VB converter was used as a starting point for creating the XML output. Instead of outputting clean VB code, the methods were rewritten to generate XML nodes. Currently this step strips out much of the information from the source code. Since the focus of this tool is not to evaluate metrics, but rather to extract traceability information, statements such as *if*, *else*, *for*, *while*, *switch*, *case*, etc. are not relevant. The variables used in these statements are recorded, but the overall structure within methods was not. Future work on this project may look at extracting metrics to evaluate not only the traceability information, but also to provide a report on the quality of the code.

With the code transformed into an XML representation, all that remains is to pull out the information that we are interested in, and then represent it in such a way that the traceability information can be visually understood. A data model was created to represent the traceability information. The XML code representation was then parsed in order to extract the information and populate the data model. Figure 1 shows the flow of information going from source code to the traceability knowledge base.

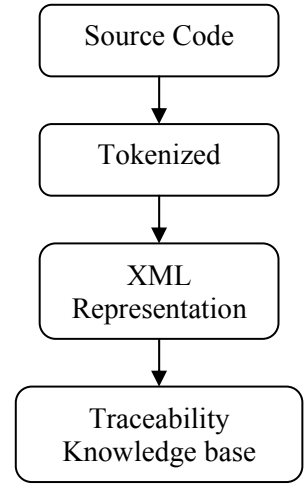

Fig. 1:    Information flow

## 5  Data Model

The data model created for this project needed to represent the traceability information for numerous different types of objects, with numerous different types of relationships. It was decided to design the data model in such a way that new categories of object types or relationships could be added to the knowledge base at any time. At the highest level, there are five major object types in the data model. The relationships between the top-level object types are shown in Figure 2. These types and their relationships are explained below.

### 5.1  KnowledgeBase Object

The topmost element in the data model, the KnowledgeBase object is essentially a wrapper that contains two lists of sub-objects. These lists contain KnowledgeType and LinkType objects, both of which are explained below. All changes made to the data model are sent to the KnowledgeBase object, which then sends out events to the listeners (the user interface mostly).

### 5.2  KnowledgeType Object

Every object that we are trying to represent in our traceability system needs to have a specific type. The current list of types developed for this system is: *Namespace*, *Class*, *Constructor*, *Method*, *Property*, *Variable*, *Delegate*, and *Event*. A KnowledgeType object represents each of these types. KnowledgeType objects point directly to all of the top level KnowledgeObjects that they contain. The user interface represents each KnowledgeType object as a column in the interface. These columns may be displayed or hidden, and can be displayed in any order. Queries to the knowledge base can be based on the KnowledgeType object.

---

[1] #Develop – www.icsharpcode.net

### 5.3  KnowledgeObject Object

The KnowledgeObject object is used to represent each object that is extracted from the source code. Every KnowledgeObject must have a specific KnowledgeType, and thus will be represented in that column in the interface. Each KnowledgeObject also contains a list of LinkObjects, which are used to represent the relationships between the KnowledgeObjects. After the automated extraction of traceability information has taken place, the user is able to add extra information to KnowledgeObjects. This information may include adding notes about important information or problems associated with the KnowledgeObject, or links to documents that describe the object.

### 5.4  LinkObject Object

All of the relationships among the different KnowledgeObject objects are represented as LinkObject objects. Each LinkObject object has a parent and child KnowledgeObject object, used to represent the directionality of the relationship. The LinkObject also specifies the LinkType of the relationship, which essentially describes the relationship between the two KnowledgeObject objects.

### 5.5  LinkType Object

As previously mentioned, every relationship in the traceability system involves two KnowledgeObject objects, which are linked together by a specific LinkObject object. Every LinkObject object has a specific LinkType object, which essentially describes the relationship between the two KnowledgeObject objects. By having these separate LinkObject and LinkType objects, the relationships between the different KnowledgeObject objects can be distinguished from each other. This allows the user to display traceability information for only certain LinkTypes if they chose.

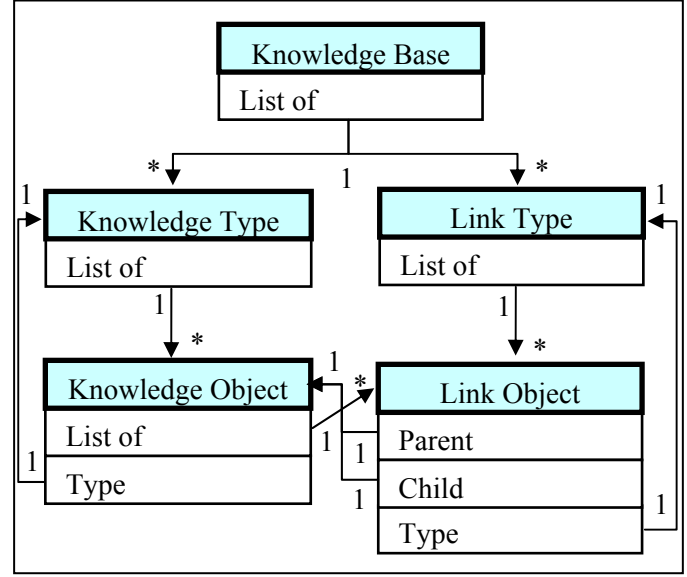

Fig. 2:    Top level objects and their relationships

## 6  User Interface

The main goal of the user interface is to represent the traceability information in a way that is easily understandable. The success of many software projects depends on the user interface. In order to clarify the different object types that are represented by the system, each object type is represented by its own column in the interface. Columns may be added or removed from the interface, as well as have their order changed. Adding a new column, for example a *Requirements* column, is as simple as creating a new KnowledgeType object in the data model. Part of the user interface is shown in Figure 3.

Every node in the different columns has a checkbox in front of it. Checking off this check box will cause the system to access the data model and determine the traceability information related to the (un)selected node. This traceability information is represented visually onscreen by parsing the lists of objects in the other columns. If a node has been checked off, then all of the objects that are related to that node are displayed in their respective column, while nodes that are not related to any of the selected nodes will not be shown. This traceability information is passed from column to column, so objects that are indirectly related to the selected object will also be displayed. If a column has no nodes checked off, then the traceability for all of the displayed nodes will be used, but once a node has been checked off, only the traceability information for the checked off nodes will be used.

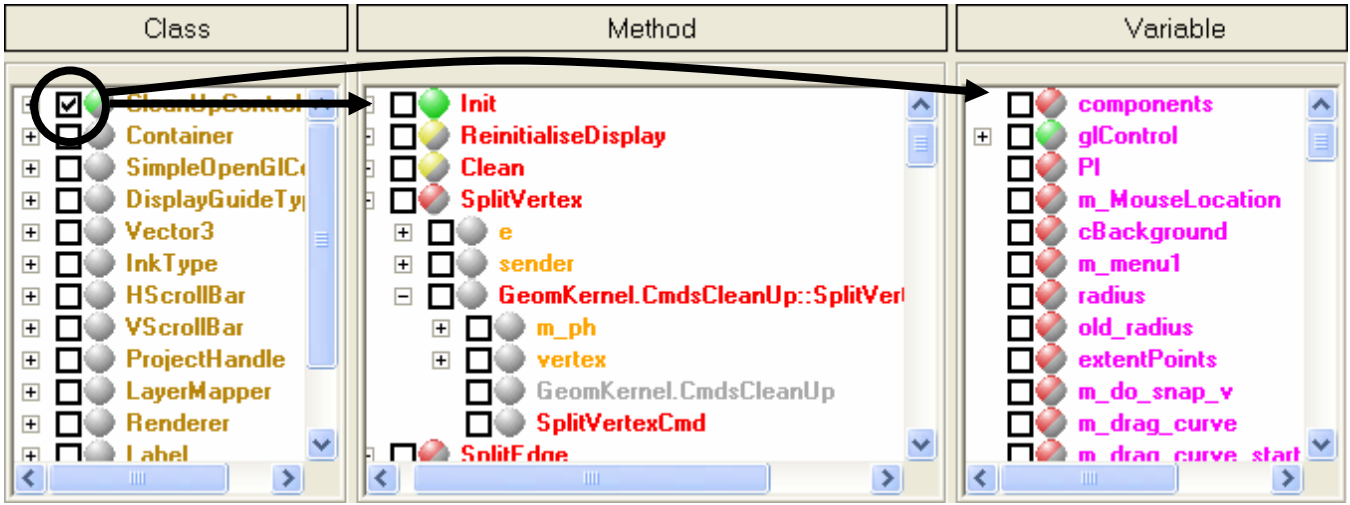

Fig. 3:    User Interface

The traceability information is also represented in tree format within the columns themselves.  The user can jump to any item of interest in the software project, and expand the node.  This will reveal all of the traceability information where the expanded node is the parent of the relationship.  Figure 4 shows a small example of what the traceability tree could look like.  In Figure 3, the Class, Method and Variable sections are shown at the top level.  The *SplitVertex* node, which has been expanded, shows how the traceability information is shown in both tree and column format for easy accessibility.

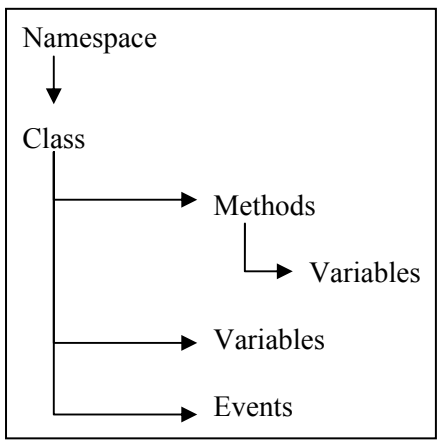

Fig. 4: Traceability Tree

Different colours of text are used to represent different types of objects.  In the example in Figure 3, red is used to denote *Methods* and *Method Calls*

(such as *Init* and *SplitVertex*), while orange is used for *Parameters* (such as *e* and *sender*) that are passed to the *Methods*.  *Variables* (such as *glControl* and *m_drag_curve*) are magenta, and *Namespaces* (such as *GeomKernel.CmdsCleanUp*) are grey.  Types not represented in Figure 3 also have distinct colours.  The coloured balls are used to represent the accessibility level of the node and its children.  Green is used for public, red is used for private, and yellow is used for all other levels.

Once the automatic extraction has populated the knowledge base with traceability information, the user is free to modify the resulting data model any way that they choose.  Simply simply dragging one object onto another object will create a relationship between them.  At the bottom of the interface, a panel exists for adding extra information about the selected object.

Each object, or node, in the different columns can be selected and information about the object will be displayed at the bottom of the control.  For the purpose of this paper, the tab control at the bottom of the interface has been expanded so that all of the tabs are visible at the same time.  There are four major tabs pages containing information.  These pages are shown in Figure 5, and their descriptions follow:

The first tab is used to display the name and description of the selected node.  The name is displayed at the top of the tab page.  Underneath the name is the object type of the selected object, and a description of the object.  This description can be generated from the comments at the object's declaration, and subsequently edited by the developers.  Version information can be used to keep track of the latest build in which this object was changed, while the ID can be generated based on the hierarchical tree of objects (namespace → class → method etc…).

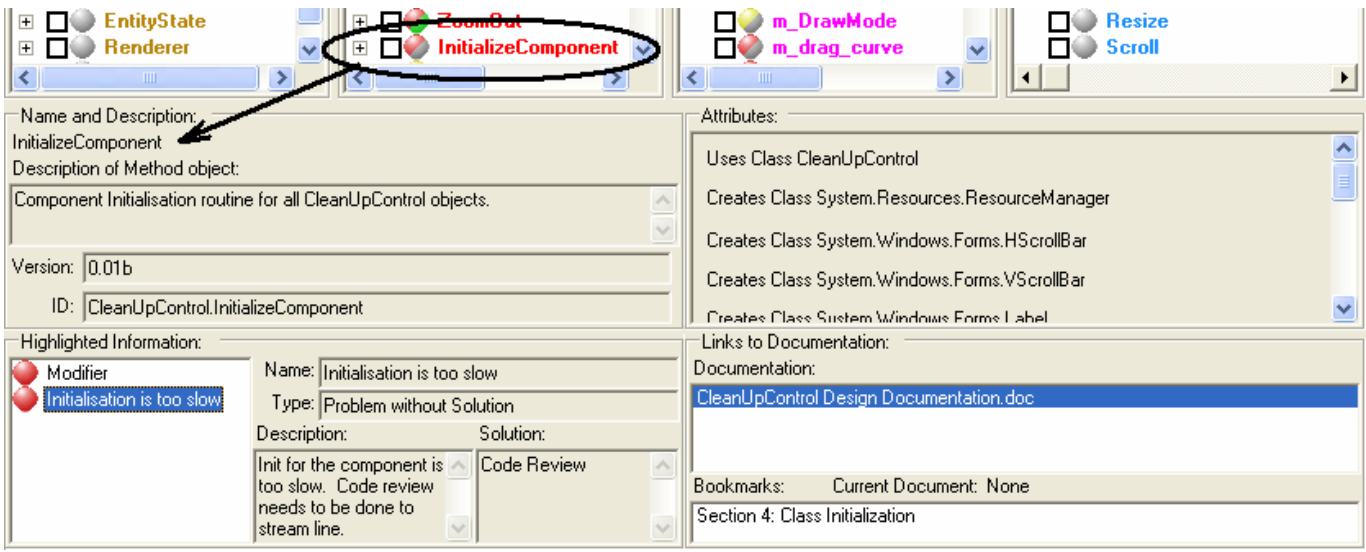

Fig. 5: Object Detail Tabs

The second tab is used to display the attributes of the selected node. These attributes can be pulled out of the code, or added by the developer through the interface. For a method, as shown in the example, the attributes list is used to list all of the different class objects that the method creates and uses as variables. Information about what methods are called by the method, as well as what objects call the method and the variables that the method uses and modifies can also be displayed in this list.

The third tab is used to add details about an object. These details can include important information that needs to be taken into consideration before changes are made, problems that have had solutions identified, and problems that have no solution currently identified. These three levels of information currently use the same system of coloured balls as the accessibility information. Future work will be to change the icons for different levels of accessibility, while keeping the colour scheme for the important information. This way, it will be easy to quickly identify the accessibility of each and every object, as well as if there is any highlighted information that should be reviewed before the object is used or updated.

The fourth tab is used to link an object to the design documentation, or any other supporting documents that the developer wishes. Any file type or web address can be linked to the object. Windows will use the default program to open a file when the user double clicks on it in the list. For Word files, bookmarks can be added to the file in order to link the relevant content to the object. Having created these bookmarks has the added benefit of warning the user when they make changes to the Word document that objects in the traceability system are linked to this portion of the document and may need to be updated as well. This warning may help to highlight what parts of code need to be updated when changes are made to the design and requirement documents.

## 7 Traceability Example

The following section presents an example of a traceability tree as it may be used by a developer or a maintainer. As mentioned previously, the different objects in the software project (namespaces, classes, methods) are presented to the user in different columns for easy navigation to the exact piece of information that the user is interested in. All of the traceability information that is found under a specific object is available from within a single column by simply expanding the object and its subsequent descendents to the level of detail that is desired.

In the example shown in Figure 6, the tree has been generated from the namespace level and expanded down to within individual methods. Important areas of the figure have been numbered so that they may be explained in further detail.

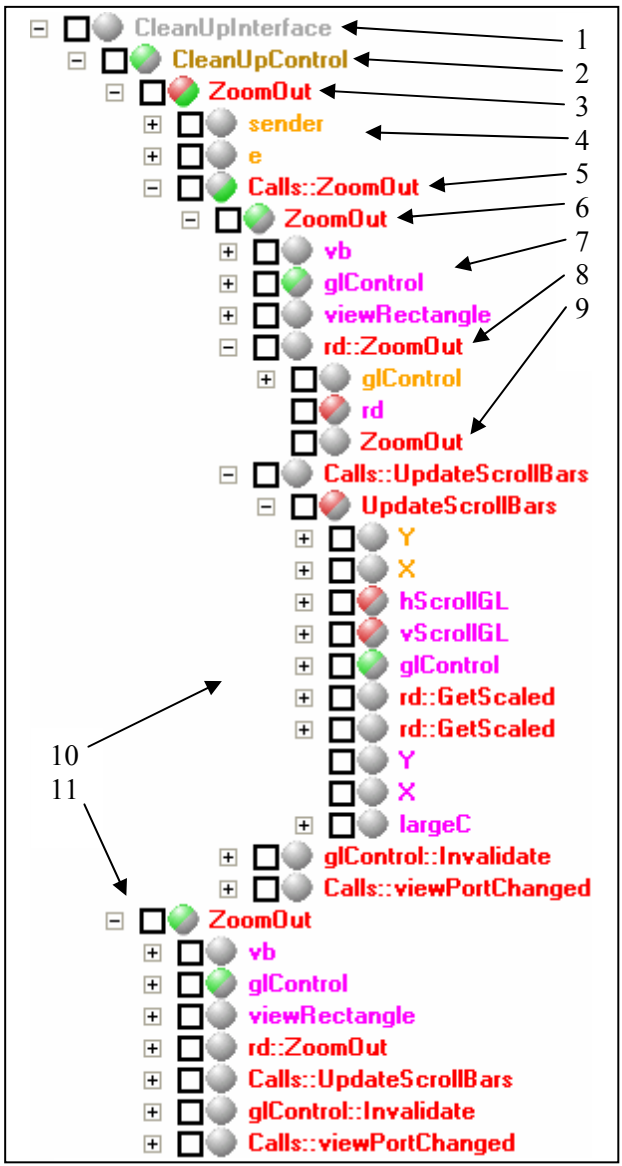

Fig. 6: Traceability Tree

1. This is the namespace object. A namespace is used to map a set of types to a common name.
2. This is a class object. A class is an object oriented concept that is used to combine data and functionality into a single entity.
3. *ZoomOut* is a method found within the *CleanUpControl* class. The arguments passed to *ZoomOut* are listed as 4.
4. The two arguments from the declaration portion of the *ZoomOut* method.
5. The *ZoomOut* method from 3 calls an overloaded version of itself. The *Calls::* indicates that this method call is to a method found within the same class.

6. The call to the overloaded method can actually be expanded to reveal the method itself. This allows the user to avoid having to navigate to other portions of the tree, which could be confusing and difficult.
7. When variables are used (either referenced or updated) they are entered into the method. This can lead to multiple entries of the same variable within a single method.
8. A method call that is called on a variable. The rd variable is an object from a class called *Renderer*, which was not loaded into the traceability system. It is known that the *glControl* is passed as an argument to the method, but the method itself (9) does not have any details since it has not been loaded into the system.
9. A method that is called by the source files used to populate the traceability database, but whose source files have not themselves been loaded. For this reason, the method can not be expanded to reveal the details.
10. Variable and Method calls are stored within a method in the order of execution. The current tool ignores looping structures and if else statements and their effects on the order of execution; however future versions will attempt to pull out this information and create sub-nodes to indicate different branches of execution. It is hoped that by doing this effectively, it will be possible to follow through the program logic as generated directly from source code, without having to read the source code itself.
11. The *ZoomOut* method which was called by an overloaded version of itself is also found as a child of the class object itself. Note that if the child nodes of 6 were collapsed, the two subnets would be exactly the same.

# 8  Future Work

This prototype has yet to be used in a case study. The first obvious future step is to have a development team use the tool on a real world project. From the team's input, it can be determined if the tool was useful, and the team's comments and suggestions can help direct the future work of this project.

As mentioned previously, future work on this project may include adding metric extraction functionality. The metrics could be represented as attributes, and visual representation could also be achieved using different colours or graphics to represent the nodes. Microsoft's Visual Studio 2005 may include the use of OOML, a mark-up schema for object oriented programming languages. If this occurs, then this tool would be able to work from the source code in real time, and perhaps be integrated into the Visual Studio IDE. Traceability information, combined with rules about objects in the source code, could flag problems for the developer before they even try to compile the code. Experience shows that real time functionality is required to gain acceptance from the development community. It is for this reason that this will be the major focus of future work.

The current major hurdle with the prototype is the level of interconnectedness between the different nodes. This interconnectedness can reduce the usability of the traceability information since for some projects virtually every method can be linked to any other method through other intermediate methods. Other points of interest include pulling out the comments from source code and attaching them to the traceability nodes, as well as finding a way to capture requirements for the project and mapping them to traceability nodes. Currently, reverse traceability is only available at a column to column level. Future versions will allow drilling down into objects to view their reverse traceability information

# 9 Conclusion

This paper has described a traceability tool developed for C# software projects. The traceability information extracted from project source code can be very useful for team members (programmers or maintainers) who are new to a software project. Generally, the documentation required to fully understand a software project is not available until the development has been completed, and even then it is often not complete. By being able to visually see the traceability information, developers are able to quickly evaluate the impact that changes they introduce will have on the entire system, as well as being able to track bugs and finding where variables are being updated by the system. Using tools such as this one and the others discussed in Section 3, the future work of maintainers should become easier, faster, and involve much less risk in a business sense. The transition times during which maintainers must learn how the software system works will continue to fall as tools such as this one are integrated into real time IDE interfaces used for development.